\documentclass[amsmath,amssymb,twocolumn]{revtex4}
\usepackage[parfill]{parskip}    % Activate to begin paragraphs with an empty line rather than an indent
\usepackage{graphicx}
\usepackage{amssymb}
\usepackage{epstopdf}
\usepackage{color}
\usepackage{graphicx}
\usepackage{pdfpages}
\begin{document}

\title{A general interior anisotropic solution for a BTZ vacuum in the context of the Minimal Geometric Deformation decoupling
approach}
\author{Ernesto Contreras {\footnote{On leave from Universidad Central de Venezuela}
\footnote{econtreras@yachaytech.edu.ec}} }
\address{School of Physical Sciences and Nanotechnology, Yachay Tech University, 100119 Urcuqu\'i, Ecuador}
\author{\'Angel Rinc\'on\footnote{arrincon@uc.cl} }
\address{Instituto de F\'isica, Pontificia Universidad Cat\'olica de Chile,\\ Av. Vicu\~na Mackenna 4860, Santiago, Chile}
\author{Pedro Bargue\~no\footnote{p.bargueno@uniandes.edu.co}}
\address{Departamento de F\'{\i}sica, Universidad de Los Andes,\\ Cra. 1 E No 18 A-10, Bogot\'a, Colombia}

\begin{abstract}
In this work we implement the Minimal Geometric Deformation decoupling method to
obtain general static interior solutions for a BTZ vacuum from the most general
isotropic solution in $2+1$ dimensions including the cosmological constant
$\Lambda$. We obtain that the general solution
can be generated only by the energy density of the original isotropic sector, so that this quantity
plays the role of a generating function. Although as a particular example we study the 
static star with constant density, the method here developed can be easily applied to more complex situations described
by other energy density profiles. 
\end{abstract}
\maketitle

\section{Introduction}\label{intro}
Obtaining new exact and interesting solutions of Einstein's field equations is 
a difficult task. In most of the cases, the
modification of certain assumptions of well known solutions, such as spherical or circularly symmetry, staticity or isotropy, in order to deal with more realistic situations ,
leads to technical difficulties to solve the equations and the use of numerical methods is usually mandatory.
However, the Minimal Geometric Deformation (MGD) method 
\cite{antoniadis1990,antoniadis1998,ovalle2008,
ovalle2009,ovalle2010,casadio2012,ovalle2013,ovalle2013a,
casadio2014,casadio2015,ovalle2015,casadio2015b,
ovalle2016, cavalcanti2016,casadio2016a,ovalle2017,
rocha2017a,rocha2017b,casadio2017a,ovalle2018,ovalle2018bis,estrada2018,ovalle2018a,lasheras2018,gabbanelli2018,
sharif2018,fernandez2018,fernandez2018b,contreras2018,estrada,contreras2018a,morales,tello18}
has become in an economic and powerful tool
to extend well known solutions of the Einstein field equations \cite{ovalle2017,ovalle2018,estrada2018,ovalle2018a,lasheras2018,gabbanelli2018,
sharif2018,contreras2018,contreras2018a,rincon2018,ovalleplb,contreras2018c,contreras2018d}. 
For example, the method has allowed to induce local anisotropies in spherically
symmetric systems leading to both more realistic interior solutions of compact objects 
\cite{lasheras2018,gabbanelli2018} and hairy black holes \cite{ovalle2018a}.  In all of these anisotropic 
extensions, the matching conditions play an important role. To be more precise, depending on 
the nature of the exterior geometry surrounding the compact object, the matching conditions lead
to particular constraints on the density or the pressure for the induced anisotropic solution. 

Up to now, several contributions have been made in the context of anisotropic relativistic stars, for instance, the pioneering work of Bowers and Liang \cite{bowers} who analysed the hydrostatic equilibrium for the case of local anisotropy obtaining a generalized equation. Since their work, anisotropic models have been extensively investigated studying the effect that local anisotropies have on the bulk properties of spherically symmetric (and static) general relativistic compact stars. 
Particular attention should be dedicated to the seminal works of Herrera et. al. about anisotropy in relativistic astrophysics. These works opened a new window in the study of anisotropic relativistic stars (see \cite{Cosenza_1981} and \cite{Cosenza_1982} and references therein).
A more realistic description of the underlying physics requires to take into account a lot of non--trivial ingredients. 
One of the main uncertainties in the description of inner solutions relies on the choice of certain equation of state, but still 
in case we know the adequate form for it, given that the Einstein field equations are highly non--linear, obtaining the corresponding physical solution could demand a lot of effort. A simple way to bypass some technical issues
is to  solve the problem at lower dimensions reason why, in
this work, we are interested in the study of the anisotropization 
of any perfect fluid solution in $2+1$ dimensions with negative cosmological constant embedded in a BTZ vacuum.\\
This work is organized as follows. In the next section we briefly review the MGD-decoupling method in $2+1$ dimensions with 
cosmological term. In section \ref{matching} we study suitable matching conditions considering BTZ as the vacuum exterior solution.
Section \ref{static} is devoted to obtain the anisotropic extension of a static perfect fluid with constant density which is finally matched to the BTZ vacuum. Some 
final comments and conclusions are left to the last section.

\section{Einstein Equations in $2+1$ space--time dimension with cosmological constant}\label{mgd}
In a recent work \cite{contreras2018d} we have considered the MGD-method in $2+1$ dimensions with 
cosmological constant. In this section we briefly review the main results obtained in \cite{contreras2018d} 
but considering a negative cosmological constant $\Lambda=-\ell^{-2}$. Finally, we shall study the matching 
conditions of these solutions embedded in a BTZ vacuum. 

Let us start by considering the Einstein field equations
\begin{eqnarray}\label{einsorig}
R_{\mu\nu}-\frac{1}{2}R g_{\mu\nu}+\frac{ g_{\mu\nu}}{\ell^{2}}=\kappa^{2}T_{\mu\nu}^{tot},
\end{eqnarray}
assuming that the total energy-momentum tensor can be written as
\begin{eqnarray}\label{total}
T_{\mu\nu}^{(tot)}=T_{\mu\nu}^{(m)}+\theta_{\mu\nu}.
\end{eqnarray}
In the above expression, $T^{\mu(m)}_{\nu}=diag(-\rho,p,p)$
stands for the energy--momentum tensor of a perfect fluid and
$\theta^{\mu}_{\nu}=diag(-\rho^{\theta},p_{r}^{\theta},p_{\perp}^{\theta})$ contain the information of the decoupler matter. 

In this work we are interested in spherically symmetric space--times so that
the line element is parametrized as
\begin{eqnarray}\label{le}
ds^{2}=-e^{\nu}dt^{2}+e^{\lambda}dr^{2}+r^{2}d\phi^{2},
\end{eqnarray}
where $\nu$ and $\lambda$ are functions of the radial coordinate $r$ only. 

Replacing  Eq. (\ref{le}) in (\ref{einsorig}), we obtain
\begin{eqnarray}
\kappa ^2 \tilde{\rho}&=&\frac{1}{\ell^{2}}+\frac{e^{-\lambda} \lambda'}{2 r}\label{eins1}\\
\kappa ^2 \tilde{p}_{r}&=&-\frac{1}{\ell^{2}}+\frac{e^{-\lambda} \nu '}{2 r}\label{eins2}\\
\kappa ^2 \tilde{p}_{\perp}&=&-\frac{1}{\ell^{2}}+\frac{1}{4} e^{-\lambda} \left(-\lambda ' \nu '+2 \nu ''+\nu '^2\right)\label{eins3}
\end{eqnarray}
where the prime denotes derivation with respect to the radial coordinate and we have defined
\begin{eqnarray}
\tilde{\rho}&=&\rho+\alpha\rho^{\theta}\label{rot}\\
\tilde{p}_{r}&=&p+\alpha p_{r}^{\theta}\label{prt}\\
\tilde{p}_{\perp}&=&p+\alpha p_{\perp}^{\theta}.\label{ppt}
\end{eqnarray} 

Following the MGD protocol, we introduce the minimal deformation
\begin{eqnarray}\label{def}
e^{-\lambda}&=&\mu +\alpha f,
\end{eqnarray}
in equations (\ref{eins1}), (\ref{eins2}) and (\ref{eins3}) to obtain
two sets of differential equations: one set describing an isotropic system sourced by
the conserved energy--momentum tensor of a perfect fluid $T^{\mu(m)}_{\nu}$
\begin{eqnarray}
\kappa ^2\rho &=& \frac{1}{\ell^{2}}-\frac{\mu '}{2 r}\label{iso1}\\
\kappa ^2 p&=& -\frac{1}{\ell^{2}}+\frac{\mu \nu '}{2  r}\label{iso2}\\
\kappa ^2 p&=& -\frac{1}{\ell^{2}} +\frac{\mu ' \nu '+\mu \left(2 \nu ''+\nu '^2\right)}{4 },\label{iso3}
\end{eqnarray}
and the other set corresponding to Einstein field equations sourced by $\theta_{\mu\nu}$ given by 
\footnote{In what follows we shall assume $\kappa^{2}=8\pi$}
\begin{eqnarray}
\kappa ^2\rho^{\theta}&=&-\frac{f'}{2 r}\label{aniso1}\\
\kappa^{2} p_{r}^{\theta}&=&\frac{f \nu '}{2 r}\label{aniso2}\\
\kappa^{2} p_{\perp}^{\theta}&=&\frac{f' \nu '+f \left(2 \nu ''+\nu '^2\right)}{4},\label{aniso3}
\end{eqnarray}

In this work, our main goal  is to extend a well known isotropic interior solution satisfying Eqs. (\ref{iso1}), (\ref{iso2}) and
(\ref{iso3}) to anisotropic domains using expressions (\ref{aniso1}), (\ref{aniso2}) and (\ref{aniso3}) and the 
minimal deformation given by Eq. (\ref{def}). In order to do so, applying appropriate matching conditions is mandatory. The next 
section is devoted to the study of matching conditions assuming that the perfect fluid is embedded in a BTZ vacuum.

\section{Matching condition}\label{matching}
In the context of compact objects, the junction conditions play a crucial role because they reveal
information about the underlying physics of the object. In particular, in 2+1 dimensional space--times, the boundary is a 
circumference and such an interface, {\it i. e.} $r=R$, limits both the inner and outer solutions.
Thus, the complete solution is divided in two parts: i) the inner solution, which is obtained using the MGD approach, and ii) 
the outer solution, which is obtained from the Einstein field equations taking as the source the geometrical 
deformation, $\theta_{\mu}^{\nu}$.
The inner solution is parametrized by the metric:
\begin{align}
\mathrm{d}s^2 &= -\text{e}^{\nu^{-}}\mathrm{d}t^2 + \text{e}^{\lambda^{-}}\mathrm{d}r^2 + r^2 \mathrm{d}\phi^2
\end{align}
where the deformation is introduced as follows:
\begin{align}\label{metric}
\text{e}^{\lambda_{-}} &= \Bigl(-\tilde{m}(r) + \left( \frac{r}{\ell} \right)^2 \Bigl)^{-1}
\end{align}
producing an effective mass according to
\begin{align}
 \tilde{m}(r) &= m(r) - \alpha f(r). 
\end{align}
It is notable that $m(r)$ is the usual inner mass and $f(r)$ is found via the MGD approach. 
Therefore, we note that the inner mass suffers a deformation after introducing the MGD formalism.
The outer solution is written as
\begin{align}
\mathrm{d}s^2 &= -\text{e}^{\nu^{+}}\mathrm{d}t^2 + \text{e}^{\lambda^{+}}\mathrm{d}r^2 + r^2 \mathrm{d}\phi^2
\end{align}
and the corresponding outer functions are obtained solving the effective Einstein field equations 
when $G_{\mu \nu} \equiv  \theta_{\mu \nu}$. In addition, the anisotropies $\theta_{\mu \nu}$ are written in terms of the classical metric functions, $\nu(r)$ and $\lambda(r)$, these functions corresponding
to that of the BTZ black hole solution.
Now, in order to connect both solutions, we use the continuity of the so--called first and second fundamental 
forms. On one hand, the first one, evaluated on the surface, is written as
\begin{align}
\Bigl[\mathrm{d}s^2 \Bigl]_{\Sigma} &= 0
\end{align}
which produces two junction conditions:
\begin{align}
\label{FF1_A}
\text{e}^{\nu^{-}} - \text{e}^{\nu^{+}} \Bigl|_{\Sigma} &=0,
\\
\label{FF1_B}
\text{e}^{-\lambda^{-}} - \text{e}^{-\lambda^{+}} \Bigl|_{\Sigma} &=0.
\end{align}
Taking advantage of \eqref{FF1_B} we can define the black hole mass via the relation
\begin{align}
-M + \left(\frac{R}{\ell}\right)^2 + \alpha f(R) &= \text{e}^{\lambda_{+}(R)} 
\end{align}
with $\tilde{m}(R) \equiv M$ and $f(R)$ the deformation at the star surface.
On the other hand, the second fundamental form is:
\begin{align} 
\Bigl[G_{\mu \nu} r^{\nu} \Bigl]_{\Sigma} &= 0
\end{align}
where $r_{\mu}$ is a radial vector. The aforementioned restriction gives
\begin{align} 
\tilde{p}_r^{-} - \tilde{p}_r^{+} \Bigl|_{\Sigma} &= 0.
\end{align}
Finally, by using the definition of $\theta_1^1$, we obtain
\begin{align}  \label{FF2}
%p^{-} - \alpha \Bigl[ (\theta_1^1)^{-} - (\theta_1^1)^{+}) \Bigl] \Bigl|%_{\Sigma} &=0 
%\\
p_{R} &= \alpha 
\left[ 
- \frac{f_R}{8 \pi} \frac{\nu'_R}{2R} 
+ \frac{g_R}{8\pi} \frac{1}{\ell^2}\left(-\tilde{m} + \left(\frac{R}{\ell}\right)^2 \right)^{-1} 
\right]
\end{align}
where we have used the definitions
\begin{align}
p_R \equiv p(R)^{-},
\hspace{0.5cm}
\nu'_R \equiv \nu'(R)^{-}.
%\hspace{0.5cm}  
%\mathcal{M} \equiv m(R)
\end{align}
In addition, it should be noticed that $g(r)$ is the geometric deformation for the outer BTZ 
solution given by the anisotropic contribution $\theta_{\mu\nu}$, where the metric has the same structure of 
Eq. (\ref{metric}) under the replacement 
\begin{align}
\text{e}^{\lambda} &= \left( -\tilde{m}  + \left(\frac{r}{\ell}\right)^2 + \alpha g(r) \right)^{-1}.
\end{align}
Equations (\ref{FF1_A}), (\ref{FF1_B}) and (\ref{FF2}) are the necessary and sufficient conditions for the matching of the 
interior MGD metric with the outer vacuum described by the deformed BTZ metric.
It is important to point out that the 2+1 dimensional case has a common feature with the four dimensional counterpart: 
if the outer region is given by a not deformed BTZ solution (which implies $g(r) = 0$) we obtain the simplest case
\begin{align}\label{preR}
\tilde{p}_{R} \equiv p_{R} + \alpha\left( \frac{f_R}{8\pi} \frac{\nu'_R}{2R}\right)  &= 0.
\end{align}
Note that by Eq. (\ref{aniso2}), 
\begin{eqnarray}
\frac{f_R}{8\pi}\frac{\nu'_R}{2R}=p_{r}^{\theta}(R)^{-}.
\end{eqnarray}
In this sense, a sufficient condition to fulfill (\ref{preR}) is to
consider the so called mimetic constrain of the pressure, namely
\begin{eqnarray}
p=p^{\theta}_{r}.
\end{eqnarray}
In what follows we shall implement the mimetic constraint to extend the isotropic 
static star with constant density in a BTZ vacuum.

\section{Static perfect fluid solutions with $\Lambda$}\label{static}
The line element of a static circularly symmetric $2+1$ dimensional space--time
parametrized as
\begin{eqnarray}\label{metricpf}
ds^{2}=-N^{2}dt^{2}+\frac{1}{G^{2}}dr^{2}+r^{2}d\phi^{2}
\end{eqnarray}
corresponds to a perfect fluid solution of the Einstein field equations whenever \cite{garcialibro}
\begin{eqnarray}
G^{2}&=&C-\Lambda r^{2}-16\pi\int\limits^{r}r\rho(r)dr\label{G}\\
N&=&n_{0}+n_{1}\int^{r}\frac{r}{G(r)}dr\label{N}
\end{eqnarray}
where $C$, $n_{0}$ and $n_{1}$ are integration constants and $\rho$ stands for the energy density. 
From the Einstein's equations, the pressure can be written as
\begin{eqnarray}\label{pr}
p=\frac{1}{8\pi N}(n_{1}G+\Lambda N).
\end{eqnarray}
Note that Eqs. (\ref{metricpf}), (\ref{G}), (\ref{N}) and (\ref{pr}) correspond to the more general 
isotropic, static and circularly symmetric solution in $2+1$ dimensions. It is worth mentioning that, after imposing 
suitable matching conditions, the above solution could serve as an interior solution for a BTZ vacuum. 

Our goal is to extend the above isotropic solution to anisotropic domains by implementing the MGD decoupling. In order to do so, 
we identify
\begin{eqnarray}
\nu&=&2\log N=2 \log \left(n_{1} \int \frac{r}{G} \, dr+n_{0}\right)\label{metricnu}\\
\mu&=& G^{2}
\end{eqnarray}
where the pressure in Eq. (\ref{pr}) can be written as
\begin{eqnarray}
p=-\frac{1}{8\pi\ell^2}+\frac{n_{1}}{8\pi(n_{1} \int \frac{r}{G} \, dr+n_{0})}.
\end{eqnarray}
In the previous expressions, the constants $C$, $n_{0}$ and $n_{1}$ must be constrained in such a manner that the pressure 
vanishes at the surface of the star, namely, $p(R)=0$.
From now on, we shall focus our attention in the implementation of the MGD method. From Eq. (\ref{aniso2}), 
the mimetic constraint, $p_{r}^{\theta}=p$ leads to
\begin{eqnarray}\label{f}
f=\frac{G}{\ell^2 n_{1}}\left(\ell^2 n_{1} G-n_{1} \int \frac{r}{G} \, dr-n_{0}\right)
\end{eqnarray}
The decoupler condition, $e^{-\lambda}=\mu+\alpha f$, leads to
\begin{eqnarray}\label{metriclambda}
e^\lambda=\frac{\ell^2 n_{1}}{G\left((\alpha +1) \ell^2 n_{1} G-
\alpha\left(n_{1} \int \frac{r}{G(r)} \, dr+n_{0}\right)\right)}.
\end{eqnarray}
Now, replacing Eq. (\ref{lambda}) in Eqs. (\ref{eins1}), (\ref{eins2}) and (\ref{eins3}) we obtain
\begin{eqnarray}
\tilde{\rho}&=&
\frac{
\alpha G' \left(n_{1} \int \frac{r}{G} \, dr+n_{0}\right)-2 (\alpha +1) \ell^2 n_{1}G'G}{16 \pi  \ell^2 n_{1} r}\nonumber\\
&&+\frac{(\alpha +2)  r}{16 \pi  \ell^2  r}\label{tilde1}\\
\tilde{p}_{r}&=&\frac{(\alpha +1)}{8 \pi } \left(-\frac{1}{\ell^2}+\frac{n_{1} G}{n_{1} \int \frac{r}{G}\, dr+n_{0}}\label{tilde2}
\right)\\
\tilde{p}_{\perp}&=&
-\frac{(\alpha +1) \left(-L^2 n_{1} G+ n_{1}\int \frac{r}{G} \, dr+n_{0}\right)}{8 \pi  L^2 \left(n_{1} \int \frac{r}{G} \, dr+n_{0}\right)}\nonumber\\
&&+\frac{\alpha  r G'}{16 \pi  L^2 G}-\frac{\alpha  n_{1} r^2}{16 \pi  L^2 G \left(n_{1} \int \frac{r}{G} \, dr+n_{0}\right)}\label{tilde3}
\end{eqnarray}
It is worth noticing that, on one hand, the system described by Eqs. (\ref{metricnu}), (\ref{metriclambda}) and
the matter content given by Eqs. (\ref{tilde1}), (\ref{tilde2}) and (\ref{tilde3}) corresponds to
a general anisotropic solution in $2+1$ dimensions with cosmological term. 
On the other hand, given the dependence of $G$ with the density (see Eq. (\ref{G})), the solution has only one generating 
function $\rho$ in the sense that, given any suitable anisotropic solution, the corresponding anisotropic system can be generated 
by Eqs. (\ref{G}), (\ref{metricnu}), (\ref{metriclambda}), (\ref{tilde1}), (\ref{tilde2}) and (\ref{tilde3}).

As a particular example we shall consider a static star with constant density $\rho_{0}$ \cite{garcialibro} which 
corresponds to
\begin{eqnarray}
e^\nu&=&\left(n_0
-\frac{n_1 \sqrt{c_1+r^2 \left(\frac{1}{\ell^2}-8 \pi  \rho _0\right)}}
{8 \pi  \rho _0-\frac{1}{\ell^2}}
\right)^{2}\\
\mu&=& c_1+r^2 \left(\frac{1}{\ell^2}-8 \pi  \rho_{0}\right),
\end{eqnarray}
with the pressure given by
\begin{align}
p=
\frac{8 \pi  \ell^4 n_1 \rho_{0} \sqrt{c_1+r^2 \left(\frac{1}{\ell^2}-8 \pi \rho_{0}\right)}+n_0 \left(1-8 \pi  \ell^2 \rho_{0}\right)}{8 \pi  \ell^2 n_0 \left(8 \pi  \ell^2 \rho_{0}-1\right)-8 \pi \ell^4 n_1 \sqrt{c_1+r^2 \left(\frac{1}{\ell^2}-8 \pi \rho_{0}\right)}}.
\end{align}
As previously stated, the constants $c_{1}$, $n_{0}$ and $n_{1}$ must be 
constrained such that $p(R)=0$. Thus, re--writing $n_0$ in terms of the other parameters 
we finally have:
\begin{align}
n_0 &= - \frac{8 \pi \rho_0}{\Lambda(\Lambda + 8 \pi \rho_0)} n_1 
\sqrt{
c_1 - (\Lambda + 8 \pi \rho_0)R^2
}
\end{align}
or, equivalently,
\begin{align}
n_0 &= \frac{8 \pi \rho_0 \ell^4}{8 \pi \rho_0 \ell^2 -1} n_1 \sqrt{c_1 + \left(\frac{1}{\ell^2} - 8 \pi \rho_0\right)R^2} .
\end{align}
From Eq. (\ref{f}), the decoupling function reads
\begin{eqnarray}
f&=&
   -8 \pi  \rho _0 r^2  
  +c_1 \left(\frac{1}{8 \pi  \ell^2 \rho _0-1}+1\right)\nonumber\\
&&  -\frac{n_0 \sqrt{c_1+r^2 \left(\frac{1}{\ell^2}-8 \pi  \rho _0\right)}}{\ell^2 n_1} 
\end{eqnarray}
Now, using (\ref{metriclambda}) we obtain
\begin{eqnarray}\label{lambda}
\lambda=
\log \left(\frac{\ell^2 n_1 \left(1-8 \pi  \ell^2 \rho_{0}\right)}{D_{\lambda}}\right)
\end{eqnarray}
where
\begin{eqnarray}
D_{\lambda}&=&-n_{1}c_{1}\ell^{2}+  \alpha  n_0 \left(8 \pi  \ell^2 \rho_{0}-1\right) \sqrt{c_1+r^2 \left(\frac{1}{\ell^2}-8 \pi  \rho_{0}\right)}\nonumber\\
&&+n_1 \left(8 \pi  (\alpha +1) \ell^2 \rho_{0}-1\right) \left(r^2 \left(8 \pi  \ell^2
\rho_{0}-1\right)\right)
\end{eqnarray}
Finally, replacing Eq. (\ref{lambda}) in Eqs. (\ref{tilde1}), (\ref{tilde2}) and (\ref{tilde3}) we obtain
\begin{align}
\tilde{\rho}&=(\alpha +1)\rho_{0}+\frac{n_0 \left(\alpha -8 \pi  \alpha  \ell^{2}\rho_{0}\right)}{16 \pi  \ell^{4} n_1 \sqrt{c_1+r^2 \left(\frac{1}{\ell^{2}}-8 \pi \rho_{0}\right)}}\\
\tilde{p}_{r}&=\frac{8 \pi  (\alpha +1) \ell^4 n_1 \rho_{0} \sqrt{c_1+r^2 \left(\frac{1}{\ell^2}-8 \pi 
\rho_{0}\right)}}{D_{pr}}\nonumber\\
&+\frac{(\alpha +1) n_0 \left(1-8 \pi  \ell^2 \rho_{0}\right)}{D_{pr}}\\
\tilde{p}_{\perp}&=-\frac{(\alpha +1) \ell^2 n_1 \rho_{0} \sqrt{c_1+r^2 \left(\frac{1}{\ell^2}-8 \pi  \rho_{0}\right)}}{\ell^2 n_1 \sqrt{c_1+r^2 \left(\frac{1}{\ell^2}-8 \pi  \rho_{0}\right)}+n_0 \left(1-8 \pi \ell^2 \rho_{0}\right)}\nonumber\\
&+\frac{(\alpha +2) n_0 r^2 \left(8 \pi  \ell^2 \rho_{0}-1\right) \left(8 \pi \ell^2 
\rho_{0}-1\right)}{D_{p\perp}}\nonumber\\
&-\frac{2 (\alpha +1) c_1 \ell^2 n_0 \left(8 \pi  \ell^2 \rho_{0}-1\right)}{D_{p\perp}}
\end{align}
where
\begin{align}
D_{pr}&=-8 \pi  \ell^4 n_1 \sqrt{c_1+r^2 \left(\frac{1}{\ell^2}-8 \pi  \rho_{0}\right)}\nonumber\\
&+8 \pi  \ell^2 n_0 \left(8 \pi  \ell^2 \rho_{0}-1\right)\\
D_{p\perp }&=16 \pi  \ell^2  \bigg(\ell^2 n_1 \sqrt{c_1+r^2 \left(\frac{1}{\ell^2}-8 \pi \rho_{0}\right)}+\nonumber\\
&n_0 \left(1-8 \pi  \ell^2 \rho_{0}\right)\bigg)\left(r^2 \left(8 \pi  \ell^2 \rho_{0}-1\right)-c_1 \ell^2\right)
\end{align}

At this point we would like to emphasize that the protocol here employed is suitable not only for the
constant density case, but for any other energy density profile. In this sense, the computations here employed pave the way for
the study of more complicated (and surely interesting) situations.

\section{Conclusions}\label{remarks}
In this work we obtained a general anisotropic solution in $2+1$ dimensions which models
a star embedded in a BTZ vacuum. The starting point was to consider the most general solution
in $2+1$ dimensions with cosmological constant, particularly those solutions
with negative $\Lambda$. Then, after applying the Minimal Geometric Deformation method and suitable matching conditions, we showed that a consistent anisotropic interior solution can be achieved 
by implementing the so called mimetic constraint of the pressure, namely, the radial pressure of the anisotropic sector is taken as the pressure of the isotropic solution. The final result is that
the general anisotropic solution depends on the energy density of the perfect fluid only, so that
this function plays the role of a generating function. 
It is worth mentioning that the method here employed
can be considered as a well--behaved protocol that can be used to construct 
anisotropic solutions in $2+1$ dimensions.
Specifically, given a suitable energy density associated to an isotropic solution, the calculation of the 
quantities describing the anisotropic solution turns to be straightforward by employing
the approach here presented.

\section*{Acknowledgments}
The author A. R. was supported by the  CONICYT-PCHA/Doctorado  Nacional/2015-21151658. 
The author P. B. was supported by the Faculty of Science and Vicerrector\'{\i}a de Investigaciones of Universidad de los Andes, Bogot\'a, Colombia.


\begin{thebibliography}{}
\bibitem{antoniadis1990} I. Antoniadis, Phys. Lett. B {\bf 246} (1990) 377.
\bibitem{antoniadis1998} I. Antoniadis, N. Arkani-Hamed, S. Dimopoulos, G. Dvali. Phys. Lett. B {\bf 436}, 257 (1998).

\bibitem{ovalle2008} J. Ovalle. Mod. Phys. Lett. A {\bf 23}, 3247 (2008).
\bibitem{ovalle2009} J. Ovalle. Int. J. Mod. Phys. D {\bf 18}, 837 (2009).
\bibitem{ovalle2010} J. Ovalle. Mod. Phys. Lett. A {\bf 25}, 3323 (2010).
\bibitem{casadio2012} R. Casadio, J. Ovalle. Phys. Lett. B {\bf 715}, 251 (2012).
\bibitem{ovalle2013} J. Ovalle, F. Linares. Phys. Rev. D {\bf 88}, 104026 (2013).
\bibitem{ovalle2013a} J. Ovalle, F. Linares, A. Pasqua, A. Sotomayor. Class. Quantum Grav. {\bf 30},175019 (2013).
\bibitem{casadio2014} R Casadio, J Ovalle, R da Rocha, Class. Quantum Grav. {\bf 31}, 045015 (2014).
\bibitem{casadio2015} R. Casadio, J. Ovalle. Class. Quantum Grav. {\bf 32}, 215020 (2015).
\bibitem{ovalle2015} J. Ovalle, L.A. Gergely, R. Casadio. Class. Quantum Grav. {\bf 32}, 045015 (2015).
\bibitem{casadio2015b} R. Casadio, J. Ovalle, R. da Rocha. EPL {\bf 110}, 40003 (2015).
\bibitem{ovalle2016} J. Ovalle. Int. J. Mod. Phys. Conf. Ser. {\bf 41}, 1660132 (2016).
\bibitem{cavalcanti2016} R. T. Cavalcanti, A. Goncalves da Silva, R. da Rocha. Class. Quantum Grav. {\bf 33}, 215007(2016).
\bibitem{casadio2016a} R. Casadio, R. da Rocha. Phys. Lett. B {\bf 763}, 434 (2016).
\bibitem{ovalle2017} J. Ovalle. Phys. Rev. D {\bf 95}, 104019 (2017).
\bibitem{rocha2017a} R. da Rocha. Phys. Rev. D {\bf 95}, 124017 (2017).

\bibitem{rocha2017b} R. da Rocha. Eur. Phys. J. C {\bf 77}, 355 (2017).
\bibitem{casadio2017a} R. Casadio, P. Nicolini, R. da Rocha. Class. Quantum Grav. {\bf 35}, 185001 (2018). 

\bibitem{ovalle2018} J. Ovalle,  R. Casadio, R. da Rocha, A. Sotomayor. Eur. Phys. J. C {\bf 78}, 122 (2018).
\bibitem{ovalle2018bis}J. Ovalle,  R. Casadio, R. da Rocha, A. Sotomayor and Z. Stuchlik, EPL {\bf 124}, 20004 (2018).
\bibitem{estrada2018} M. Estrada, F. Tello-Ortiz. Eur. Phys. J. Plus {\bf 133},  453 (2018) .
\bibitem{ovalle2018a} J. Ovalle, R. Casadio, R. da Rocha, A. Sotomayor, Z. Stuchlik,
 Eur. Phys. J. C {\bf 78}, 960 (2018).
\bibitem{lasheras2018} C. Las Heras, P. Leon. Fortschr. Phys. {\bf 66}, 1800036 (2018).
\bibitem{gabbanelli2018} L. Gabbanelli,  A. Rinc\'on, C. Rubio. Eur. Phys. J. C {\bf 78}  370 (2018).
\bibitem{sharif2018} M. Sharif, Sobia Sadiq,  Eur. Phys. J. C {\bf 78}, 410 (2018).
\bibitem{fernandez2018} A. Fernandes-Silva, A. J. Ferreira-Martins, R. da Rocha. 
Eur. Phys. J. C {\bf 78}, 631 (2018).

\bibitem{fernandez2018b}A. Fernandes-Silva, R. da Rocha. Eur. Phys.J. C {\bf 78}, 271 (2018) .

\bibitem{contreras2018} E. Contreras and P. Bargue\~no. Eur. Phys. J. C {\bf 78}, 558 (2018). 

\bibitem{morales} E. Morales, F. Tello-Ortiz,  Eur. Phys. J. C{\bf 78}, 841 (2018).  
\bibitem{tello18} E. Morales, F. Tello-Ortiz, Eur. Phys. J. C {\bf 78}, 618 (2018).

\bibitem{contreras2018a} E. Contreras, Eur. Phys. J. C {\bf 78}, 678 (2018).

\bibitem{rincon2018} G. Panotopoulos, \'A. Rinc\'on, Eur. Phys. J. C {\bf 78}, 851 (2018)

\bibitem{ovalleplb} J. Ovalle, Phys. Lett. B {\bf 788}, 213 (2019). 

\bibitem{contreras2018c} E. Contreras and P. Bargue\~no, Eur. Phys. J. C {\bf 78}, 985 (2018).

\bibitem{estrada}M. Estrada, R. Prado, arXiv:1809.03591.

\bibitem{contreras2018d} E. Contreras, arXiv:1901.00231.

\bibitem{garcialibro} A. Garc\'ia--D\'iaz. Exact solutions in three dimensional gravity, Cambridge University Press, 2017.

\bibitem{bowers} R. L.~ Bowers and E. P. T.~ Liang, \emph{ Astrophys. J.}  \textbf{188} 657, (1974).

\bibitem{Cosenza_1981}
Cosenza, M., Herrera, L., Esculpi, M.,  Witten, L. (1981). Some models of anisotropic spheres in general relativity. Journal of Mathematical Physics, 22(1), 118-125.

\bibitem{Cosenza_1982}
Cosenza, M., Herrera, L., Esculpi, M.,  Witten, L. (1982). Evolution of radiating anisotropic spheres in general relativity. Physical Review D, 25(10), 2527.

\end{thebibliography}
\end{document}